\documentstyle[11pt,newpasp,twoside,psfig,epsf]{article}
\def\edcomment#1{\iffalse\marginpar{\raggedright\sl#1\/}\else\relax\fi}
\reversemarginpar

\begin{document}
\title{The Heating of the Intra Cluster Medium} 
\author{Paolo Tozzi}
\affil{Oss. Astronomico di Trieste, via G.B. Tiepolo 11, 34100
Trieste -- Italy}

% \author{Ima Co-Author}
% \affil{The Name of My Institution, The Full Address of My Institution}

\begin{abstract}
X--ray observations indicate that non--gravitational processes play a
key role in determining the distribution of the diffuse, X-ray
emitting gas in clusters of galaxies (ICM).  The effect of
non--gravitational processes is imprinted in the ICM as an entropy
minimum.  Preheating models assume that the entropy minimum is present
in the cosmic baryons well before collapse.  On the other hand,
observations of baryons in Ly$_\alpha$ clouds show only a modest extra
heating, ruling out the presence of such an entropy plateau in the
majority of low--density baryons at high $z$.  The problem is avoided
in the {\sl internal heating} scenario, where the heating occurs only
inside virialized structures.  However, for internal heating the
energy needed to build the entropy minimum is in excess of 1 keV per
particle.  It is not clear which kind of source can heat the baryons:
SNae seem to be inefficient by a factor of $3$ or more.  This energy
crisis must be solved by other sources (like AGNs), unless we are
missing some key aspect of the heating mechanism.  The main questions
for the next years will be: when and where is the excess entropy
produced, and by which mechanism?
\end{abstract}

\section{Introduction}
Observations in the X--ray band provided many evidences for
non--gravitational heating of the diffuse, high density baryons in the
potential wells of groups and clusters of galaxies (Intra Cluster
Medium, or ICM).  The first evidence is the shape of the L--T
relation, which is steeper than the self--similar behaviour $L\propto
T^2$ predicted in the case of gravitational processes only.  Recently,
Ponman, Cannon \& Navarro (1999, PCN; see also panel B in Figure 1)
found directly an entropy excess with respect to the level expected
from gravitational heating in the center of groups.  The entropy is
defined as $S\equiv \log K$, where $K = kT/\mu m_p \rho_e^{2/3}$, $kT$
is the temperature, $\rho_e$ is the mass density of the ionized
plasma, and $\mu$ is the mean molecular weight (see also Balogh, Babul
\& Patton 1999).  The excess entropy induces a larger pressure support
that decreases the density in the central regions.  This, in turn,
rapidly decreases the X--ray luminosity which is proportional to the
square of the density.  The effect is stronger in small groups, where
the energy responsible for the entropy is comparable to the
gravitational one, while clusters, where gravity is dominant, are
mostly unaffected.  This produces a steepening of the $L$--$T$
relation.  Non--gravitational heating of the ICM is expected also on
the basis of observations of an average metallicity $Z\simeq 0.3 ~
Z_\odot$.  In fact, SNae are likely to be responsible of heating and
polluting the ICM at the same time.  For example, Finoguenov, Arnaud
\& David (2000) derived about 1 keV per particle from SNae from the
abundance of Silicon.  It should be understood how much of this energy
goes into the ICM and if it is enough to build the observed entropy
plateau.

\begin{figure}
\plottwo{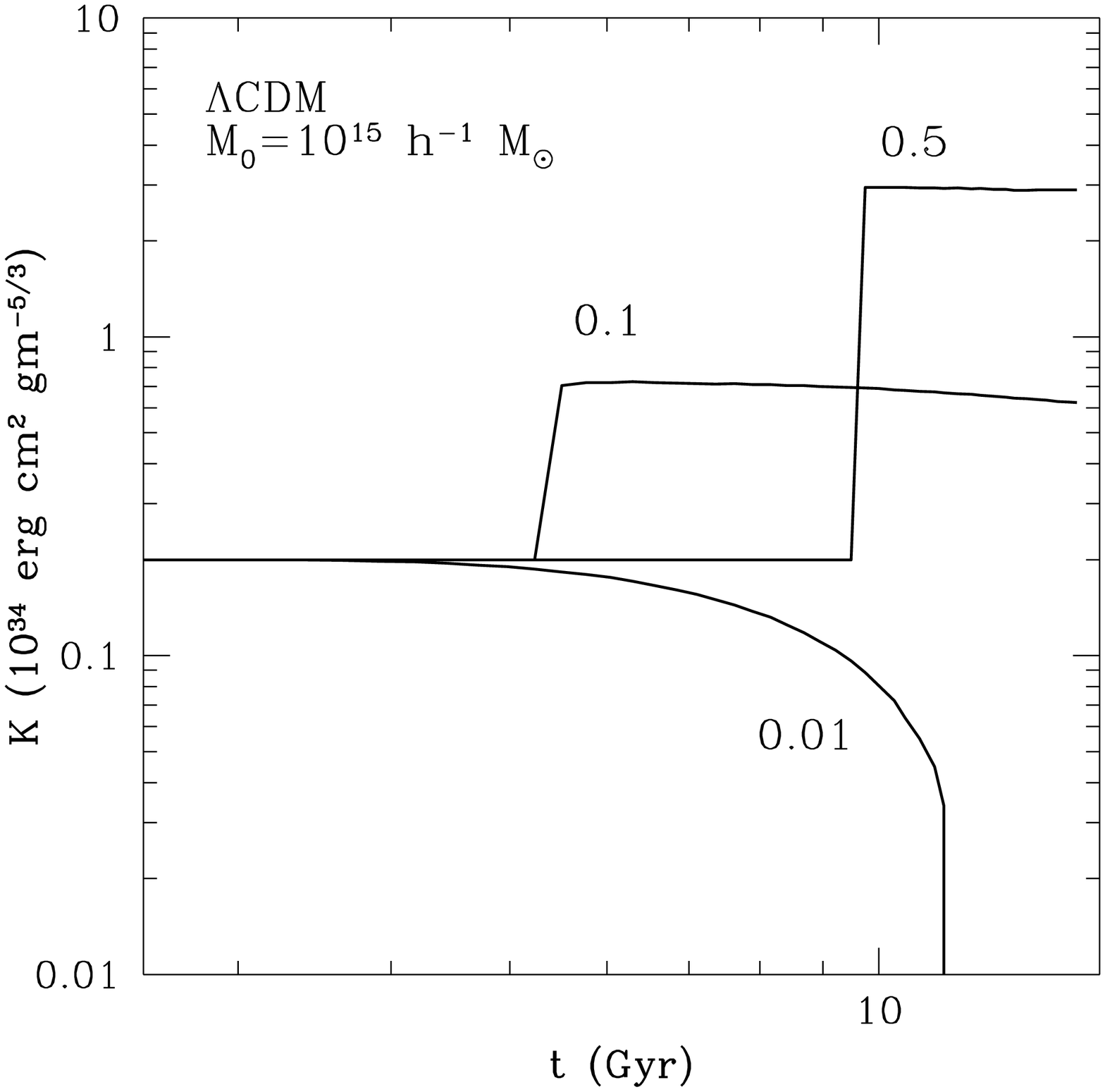}{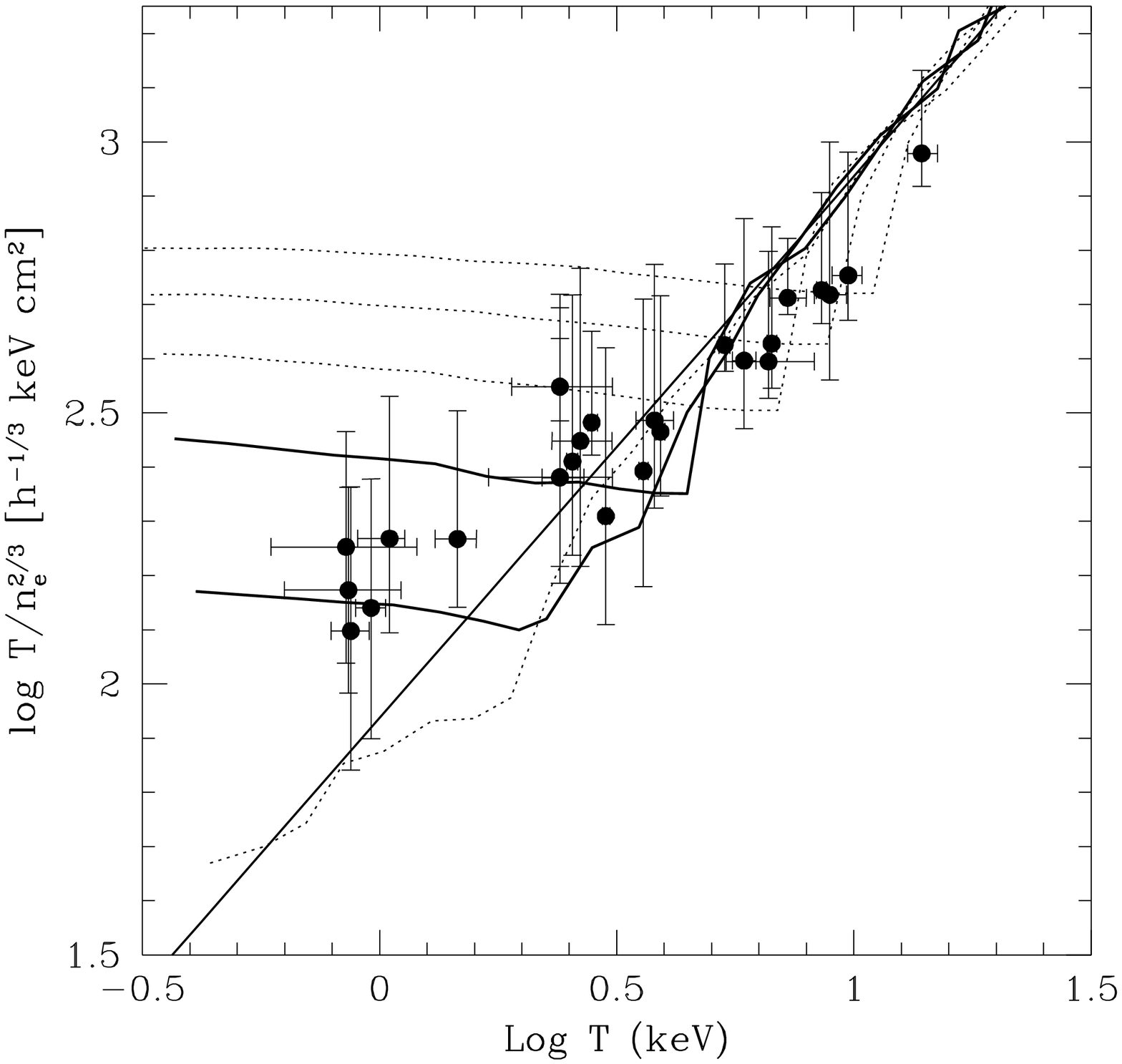}
\caption{Left: The evolution of the adiabat $K$ for three baryonic
shells (including 1\%, 20\% and 50\% of the total baryons at $z=0$) as
a function of cosmic epoch $t$ ($\Lambda$CDM cosmology, for a final
mass of $10^{15} \, h^{-1}\, M_\odot$, $K_* = 0.3 \times 10^{34}$ erg
cm$^{2}$ g$^{-5/3}$).  The sharp discontinuity occurs at the shock
(TN). Right: the relation between the entropy at $r=0.1 R_v$ and the
emission--weighted temperature for $\Lambda$CDM at redshift $z=0$ for
different values of the excess entropy $K_* = 0.6 - 0.5 - 0.4 - 0.3 -
0.2 - 0.1 \times 10^{34}$ erg cm$^2$ g$^{-5/3}$, from top to
bottom (TN). The straight line is the self similar case from N--body
simulations (see PCN).
\label{fig1}}
\end{figure}

\section{The entropy distribution of the ICM}

A first step to understand the role of entropy is to investigate its
effect on the X--ray properties of groups and clusters of galaxies.
The simplest choice is that of the {\sl external heating} models (or
preheating, see Tozzi \& Norman 2001, hereafter TN).  If the excess
entropy is present in the baryons before collapse, it will be
preserved in the cores of dark matter halos after virialization.  In
virtue of the extra pressure support, in fact, the gas is accreted
adiabatically without shock heating.  The entropy has also the effect
of suppressing the radiative cooling in the central regions (due to
the low density).  The evolution of the adiabat $K$ for the accreted
shells is shown in Figure 1 (left).  The pre--collapse entropy $K_*$
is preserved in the core of groups, in agreement with observations
(Figure 1, right).

This model allows to trace the evolution of X--ray luminosity and
temperature of groups and clusters after assuming a Press \&
Schechter--like law for the accretion rate of baryons (see TN).  In
Figure 2 we show that the evolution corresponds to tracks moving along
the local $L$--$T$ relation.  Thus, a constant $L$--$T$ is predicted
up to $z\simeq 1$, in agreement with observations (Mushotzky \& Scharf
1997, Borgani et al. 2001).  The entropy level that satisfies the
observations is between $K_* = 0.2 - 0.3 \times 10^{34}$ erg cm$^2$
g$^{-5/3}$ (see Figure 1 and 2).

\begin{figure}
\plottwo{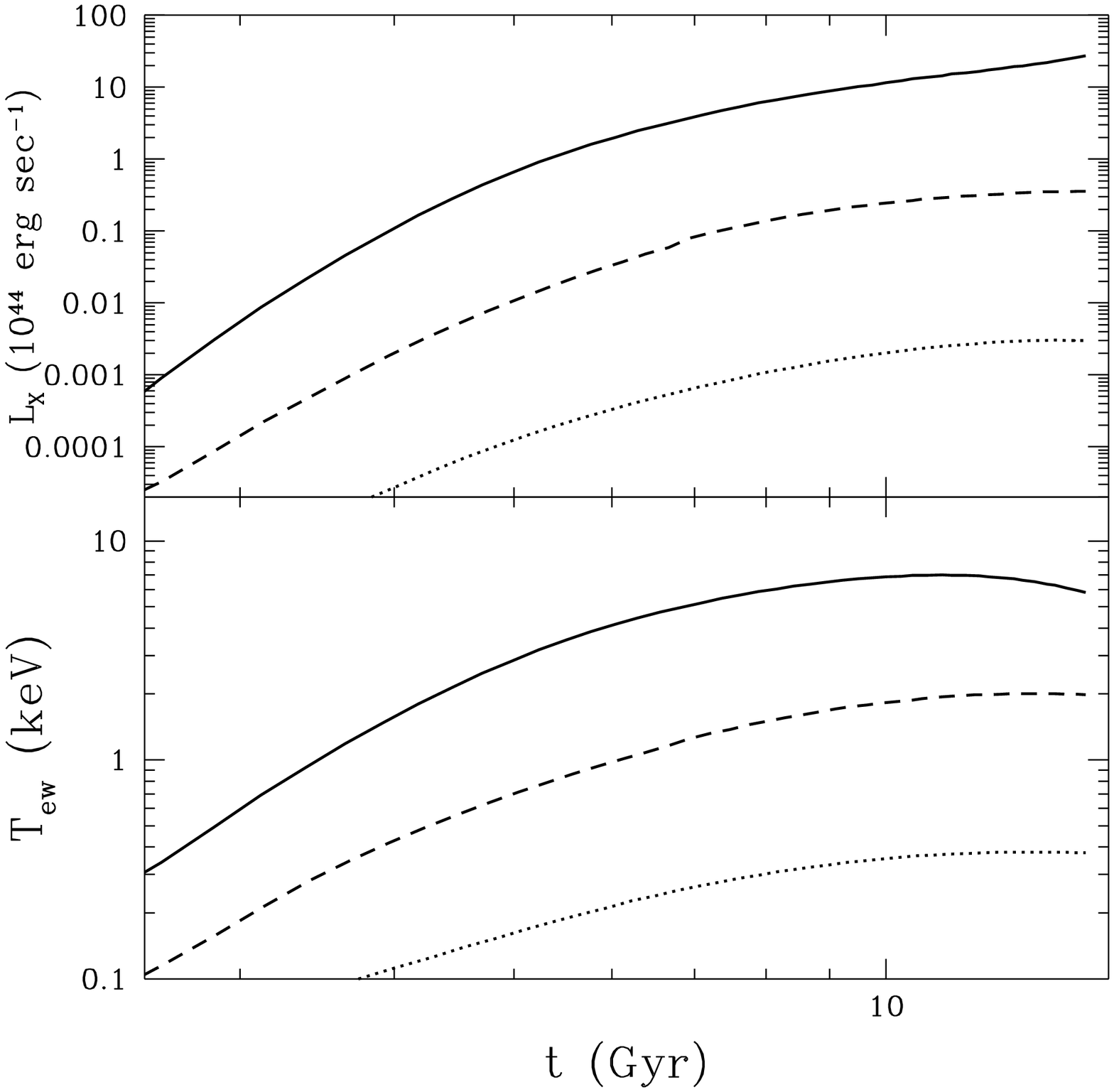}{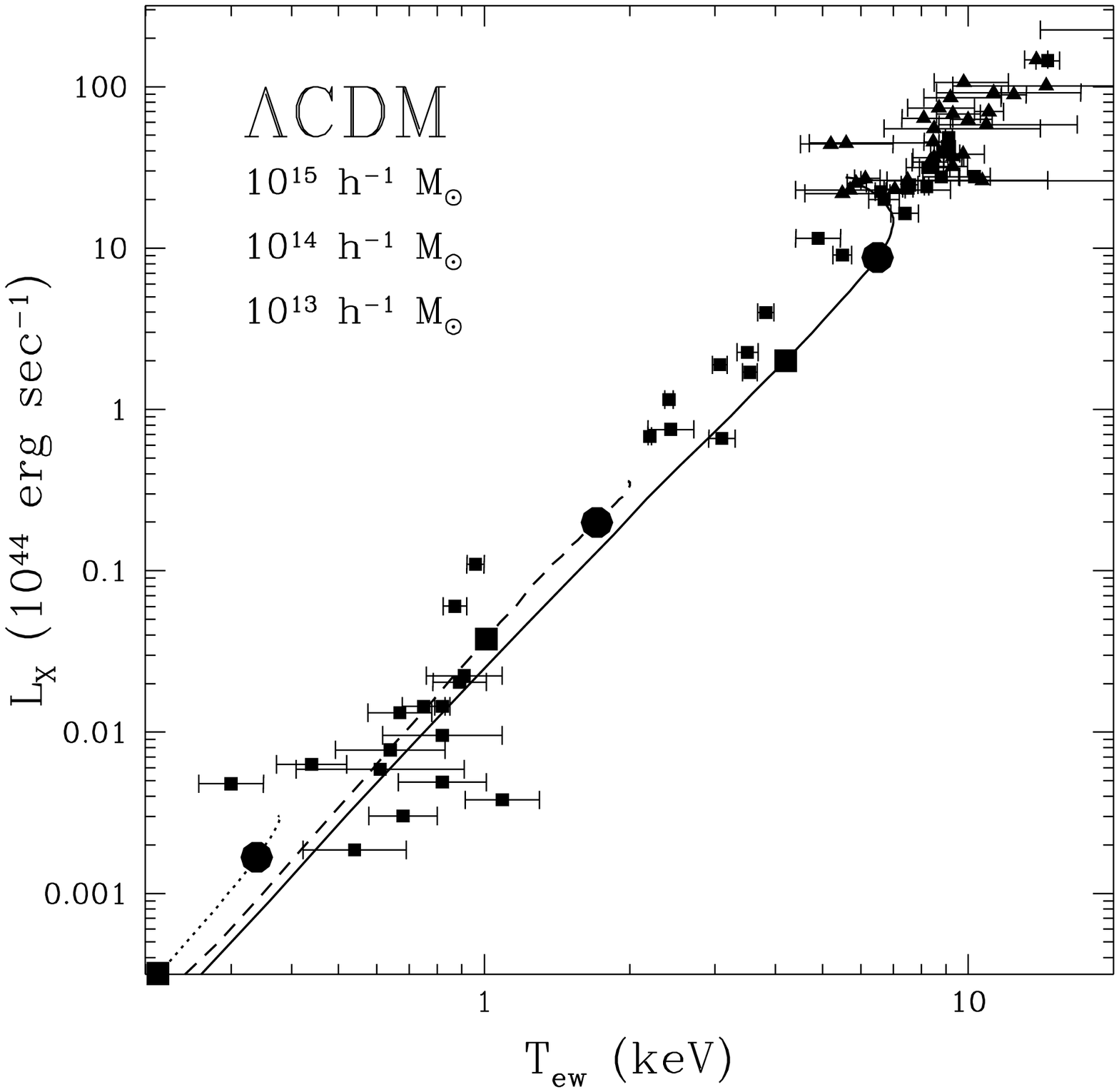}
\caption{Left: The evolution of the bolometric luminosity $L_X$
and of the emission--weighted temperature $T_{ew}$ is shown as a
function of cosmic epoch for a final mass of $10^{15} - 10^{14} -
10^{13} h^{-1} M_\odot$ (solid, dashed and dotted lines respectively)
for a $\Lambda$CDM cosmology, with a constant initial adiabat $K_* =
0.3 \times 10^{34}$ erg cm$^{2}$ g$^{-5/3}$.  Right: The evolutionary
tracks along the local $L$--$T$ relation for the halos on the left.  The
large squares and circles mark redshifts 1 and 0.5 respectively.  Data
from Allen \& Fabian (1998), Arnaud \& Evrard (1999), Ponman et
al. (1996).}
\label{fig2}
\end{figure}

In Figure 3 we show surface brightness and temperature profile for a
rich cluster and a group in the external entropy scenario (from Tozzi,
Scharf \& Norman 2001, hereafter TSN).  The treatment in TSN allows to
trace the density profile at radii larger than the virial one, in
order to check whether it is possible to detect the infalling gas.
The first case is the most favourable from the point of view of the
observations, since the very high external temperature make the
infalling gas detectable in emission around rich clusters.  This case
is expected if the external gas reached a very high entropy at
redshift zero or if it reached large overdensities due to infalling
substructures (see TSN). Inside the shock radius but outside the
adiabatic core, the entropy profile is well approximated by $K\propto
R^{1.1}$ (see Figure 4, solid lines), a behaviour which is also found
in N--body simulations (see S. Borgani, these Proceedings).

The second case is for a group with a constant external entropy $K_*
=0.3 \times 10^{34}$ erg cm$^2$ g$^{-5/3}$.  At this mass scale, the
external entropy inhibits shock heating and the accretion is entirely
adiabatic.  The entropy profile is flat and there is no discontinuity
clearly separating the accreted gas from the infalling, external gas.
In this case a substantial temperature gradient is expected in the
outer regions.  So far, the temperature profile for groups has been
observed only in the very central regions, where no significant
gradients are predicted.

\begin{figure}
\centerline{\psfig{figure=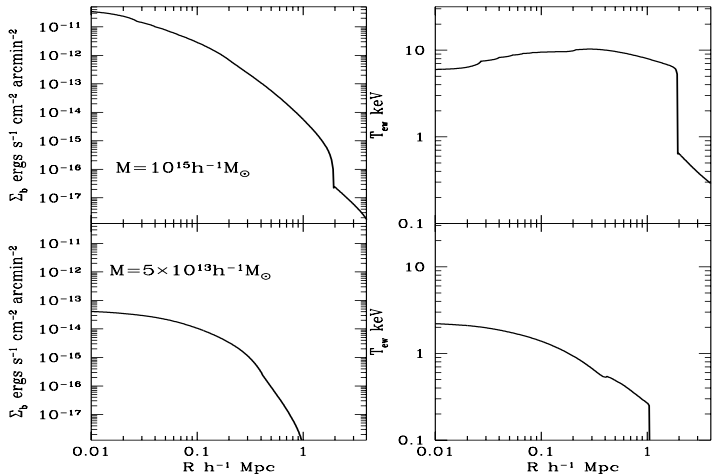,height=7cm,width=10cm}}
\caption{Surface brightness and projected temperature profiles for a
cluster of $M=1.4 ~\times ~ 10^{15} h^{-1} M_\odot$, tCDM, $K_*=3
\times 10^{34} (1+z)^{-2}$ erg cm$^2$ g$^{-5/3}$ (top panels) and $M=5
~ \times ~ 10^{13} h^{-1} M_\odot$, $\Lambda$CDM, $K_*=0.3\times
10^{34} $ erg cm$^2$ g$^{-5/3}$ (lower panels), from TSN.}
\label{fig3}
\end{figure}

\section{The energy crisis}

The external heating scenario is appealing, but it has a problem: the
excess entropy cannot be spread uniformly into the cosmic baryons at
high redshifts since this would make the Ly$_\alpha$ forest to
disappear.  Recently, it has been estimated that the level of
preheating in the Ly$_\alpha$ clouds is of the order of few $10^4 K$
(Cen \& Brian 2001), corresponding to an entropy level more than one
order of magnitude lower than that observed in groups.  A possible
solution is that only the baryons that end up in the cores of groups
and clusters are heated by a biased distribution of sources.  Such a
warm, low density gas would be unobservable at high $z$, and can be
detected as OVI absorption systems at low (Tripp, Savage \& Jenkins
2001) or at intermediate $z$ (Reimers et al. 2001).

Another solution is that the ICM is heated after the collapse.  Of
course, for a given entropy level, the much higher density implies a
much higher energy input.  An energy budget of 1--2 keV per particle
seems to be required to reproduce the entropy floor (see also Wu,
Fabian \& Nulsen 2000, Valageas \& Silk 1999) and entropy profiles
very similar to the ones predicted in the external heating scenario
(see Figure 4, dashed lines).  We recall that the preheating scenario
requires few tenths of keV (with a strict lower limit of 0.1 keV, TN)
since the gas is heated at about the background density.  Therefore,
the internal scenario requires an energy input 3--10 times higher than
that in the external scenario.

\begin{figure}
\centerline{
\psfig{figure=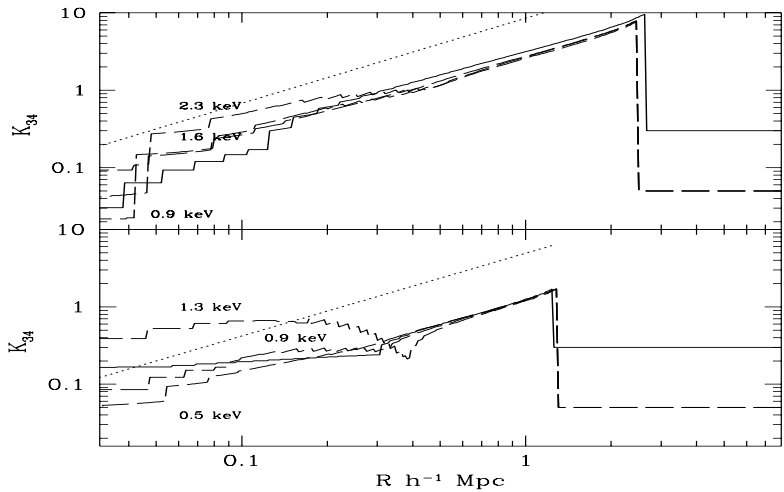,height=7cm,width=11cm}}
\caption{The entropy profiles in the external (continuous lines) and
the internal (dashed lines) scenario for a rich cluster ($M=10^{15}
h^{-1} M_\odot$, upper panel) and a small cluster ($M=10^{14} h^{-1}
M_\odot$, lower panel).  The entropy level is $K_{34} = 0.3$ (in units
of $10^{34} $ erg cm$^2$ g$^{-5/3}$) in the external scenario.  In the
internal scenario, a total energy of 1--2 keV per particle is
released, as shown by the labels.  The dotted line is the power law
$K\propto R^{1.1}$ (see TSN).}
\label{fig5}
\end{figure}

Despite the SNae can provide a large amount of energy, their
efficiency in heating the gas is unknown.  In particular, if the
heated gas has high density, the thermal energy received from SNae is
easily radiated away, with a small net increase in the gas entropy.
Simulations with a self--consistent SNae heating model plugged in,
tell us that they can hardly increase the entropy to the observed
level in groups and clusters (see talks by S. Borgani).  However we
still do not have a complete scenario to follow the heating and
enrichment by SNae and the consequent entropy history of the
surrounding baryons.

\begin{figure}
\centerline{
\psfig{figure=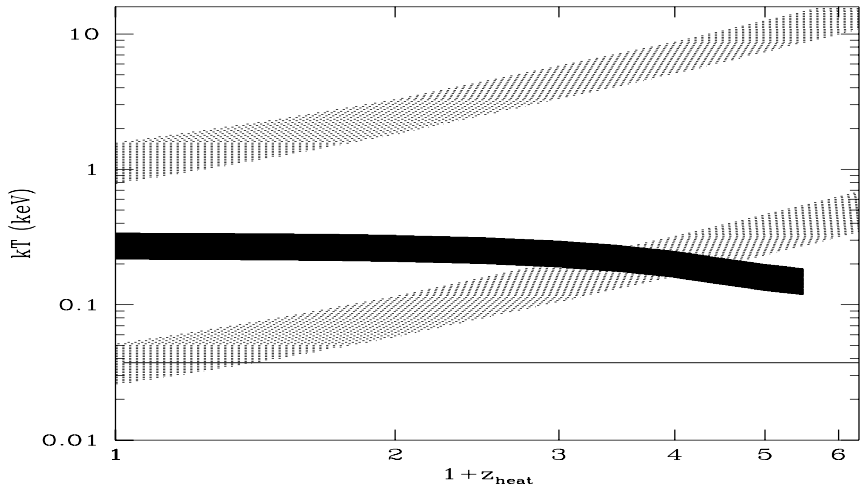,height=7cm,width=12cm}}
\caption{The upper and lower stripes show the required energy per
particle needed to obtain an excess entropy of $K_* = 0.2 - 0.3 \times
10^{34}$ erg cm$^2$ g$^{-5/3}$ in virialized structures and in the
diffuse baryons at the background density respectively.  The solid
stripe is the energy per particle dumped in the ICM by Type II and
TypeIa SNae after the calculation of Pipino et al. (2001).  The solid
line is the upper limit for the extra energy present in Ly$_\alpha$
clouds (Cen \& Bryan 2001).  }
\label{fig5}
\end{figure}

A recent progress in this direction has been made by Pipino et
al. (these Proceedings).  They computed metals and energy dumped in
the ICM starting from the observed luminosity function of cluster
galaxies.  The efficiency of TypeII and TypeIa SNae is computed
assuming a spherical gas distribution around galaxies.  The energy per
particle dumped in the ICM as a function of redshift is plotted in
Figure 5 (black stripe).  We also plotted the energy required to
obtain the entropy $K_* = 0.2 - 0.3 \times 10^{34}$ erg cm$^2$
g$^{-5/3}$ at a given redshift in virialized structures (upper stripe)
and in the background baryons (lower stripe).  It turns out that SNae
can contribute a substantial amount of the required energy.  To
understand in detail their role in the entropy history of the ICM, we
must run detailed hydrodynamical simulations (see talks by R. Bower
and A. Ferrara) to follow the heating along with the accretion of the
diffuse baryons and the cooling of the low entropy gas.  If SNae will
be shown to be unadequate for this job, it will be necessary to look
for other sources, like AGNs, which in principle can provide the
largest amount of energies (but see Yamada \& Fujita 2001).  Both
X--ray observations and theoretical modelling will be crucial in the
next few years.

\end{document}